# Existence of Oscillations in Cyclic Gene Regulatory Networks with Time Delay [*]

Masaaki Takada, Yutaka Hori, and Shinji Hara [†]

## Abstract

This paper is concerned with conditions for the existence of oscillations in gene regulatory networks with negative cyclic feedback, where time delays in transcription, translation and translocation process are explicitly considered. The primary goal of this paper is to propose systematic analysis tools that are useful for a broad class of cyclic gene regulatory networks, and to provide novel biological insights. To this end, we adopt a simplified model that is suitable for capturing the essence of a large class of gene regulatory networks. It is first shown that local instability of the unique equilibrium state results in oscillations based on a Poincaré-Bendixson type theorem. Then, a graphical existence condition, which is equivalent to the local instability of a unique equilibrium, is derived. Based on the graphical condition, the existence condition is analytically presented in terms of biochemical parameters. This allows us to find the dimensionless parameters that primarily affect the existence of oscillations, and to provide biological insights. The analytic conditions and biological insights are illustrated with two existing biochemical networks, Repressilator and the Hes7 gene regulatory networks.

## 1 Introduction

Periodic bodily functions are often related with oscillatory gene expression in living cells. In order to understand the underlying mechanism of the complex oscillatory dynamics, many theoretical works have been devoted over the past decades (see Klipp et al. (2009) for example). Most of the theoretical studies rely on the numerical simulations of the detailed mathematical models of specific biological systems. However, such approaches hardly provide a unified insight that is applicable to a broad class of biological networks. To overcome these challenges, we here consider a simplified model that is suitable for capturing the essence of a large class of gene regulatory networks, and develop analytic tools that are useful for systematically studying the existence of oscillations.

One of the pioneering theoretical analyses of oscillatory gene expression was presented in Goodwin (1965), where the dynamical model of cyclically interconnected gene's products was introduced. Later, the cyclic feedback structure was found in metabolic and cellular signaling pathways as well (Kholodenko, 2000; Stephanopoulos et al., 1998), and it is recently considered that cyclic structure plays a key role to produce the various dynamical behaviors of protein levels (see Hori et al. (2011) and references therein). In fact, the artificially constructed biological oscillator, named Repressilator (Elowitz and Leibler, 2000), was performed with a simple cyclic interaction of repressors in *Escherichia coli*. These works suggest that the study of cyclic gene regulatory networks is the first important step toward the comprehensive understanding of the large-scale gene regulatory networks in nature.

---

[*]Corresponding author Y. Hori. Tel. +81-3-5841-6893. Fax +81-3-5841-7961.

[†]M. Takada is with Department of Mathematical Engineering and Information Physics, The University of Tokyo, 7-3-1 Hongo, Bunkyo-ku, Tokyo 113-8656, Japan. takada@sat.t.u-tokyo.ac.jp   Y. Hori and S. Hara are with Department of Information Physics and Computing, The University of Tokyo, 7-3-1 Hongo, Bunkyo-ku, Tokyo 113-8656, Japan. {Yutaka_Hori, Shinji_Hara}@ipc.i.u-tokyo.ac.jp

The dynamical properties of cyclic gene regulatory networks have been actively studied in recent years, ranging from stability (Arcak and Sontag, 2006, 2008; Thron, 1991) to oscillations (El-Samad et al., 2005; Hori et al., 2011). In El-Samad et al. (2005), the parameter conditions for the existence of oscillations were considered for Repressilator, and these conditions were generalized to a general cyclic gene regulatory networks in Hori et al. (2011). A remarkable feature of these results is that the conditions are analytically obtained in terms of biochemical parameters despite the nonlinearity of the system. In particular, the dependence of the equilibrium point on the system's parameters is explicitly analyzed. Thus, one can easily see the relation between the parameters and the existence of oscillations for a broad class of genetic networks. As a result, novel essential parameters that primarily determine the existence of oscillations were found in Hori et al. (2011).

In these previous works, however, the inherent time delays in transcription, translation and translocation process were not considered in the model. Such time delays are essential especially for eukaryotic cells, because mRNA and protein productions occur at different locations in a cell, and the transportation of these substances results in sizable time delays (Chen and Aihara, 2002). The existence of oscillations was studied for the gene regulatory networks with time delay in Chen and Aihara (2002) and Enciso (2000). In these papers, however, the relation between the parameters and the existence of oscillations was not obtained in an explicit way.

The objective of this paper is (i) to provide an analytic framework to study the existence of oscillations in cyclic gene regulatory networks with time delay, and (ii) to characterize the condition for the existence of oscillations with essential parameters. To this end, we first show that local instability of the unique equilibrium state results in oscillations of protein concentrations based on the Poincaré-Bendixson theorem for cyclic time delay systems (Mallet-Paret and Sell, 1996). This reduces the analysis to the local stability analysis of a unique equilibrium of the networked time delay systems. The main theoretical contribution of this paper is the derivation of graphical and analytic conditions for the existence of oscillations. In particular, the dependence of the equilibrium on the system's parameters is explicitly considered, thus the analytic condition is obtained only in terms of biochemical parameters. These results are demonstrated with two existing gene regulatory networks, Repressilator (Elowitz and Leibler, 2000) and the somitogenesis clock (Hirata et al., 2004). We also present the effect of time delays on oscillations based on the analytic conditions.

This paper is organized as follows. In Section 2, the model of cyclic gene regulatory networks with time delay is introduced, and its dynamical properties are considered from the viewpoint of nonlinear analysis. Then, the linearized model is systematically constructed in Section 3. In Section 4, we derive the graphical condition for the existence of oscillations. Based on this result, Section 5 provides the analytic conditions. In Section 6, a toy numerical example is demonstrated to elucidate the developed conditions, and biological insights are also presented. Section 7 is devoted to the analysis of two existing biological networks. Finally, concluding remarks are given in Section 8.

A part of the results in this paper was previously presented in the authors' conference paper (Takada et al., 2010) with the omission of some details. In this version of the paper, we include the complete proofs of the theorems and the detailed description of the numerical simulations. Furthermore, the relation of our developed theorem and a conflicting theoretical result (Chen and Aihara, 2002) is clarified. Specifically, we disprove Theorem 2 in Chen and Aihara (2002), and illustrate with a counter example. It is also the first time to present the analysis result of the somitogenesis clock in Section 7.

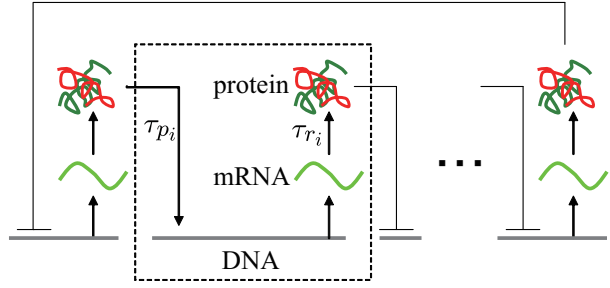

Figure 1: Cyclic gene regulatory network with time delay

# 2 Modeling and Nonlinear Analysis for Gene Regulatory Networks with Time Delay

## 2.1 Model of cyclic gene regulatory network with time delay

The well-known central dogma of molecular biology states that gene expression consists of the transcription and translation steps. During the transcription step, genes are decoded into molecules called messenger RNA (mRNA). Then, the information coded in mRNA is translated into proteins during the translation step. The rate of mRNA production is affected by the proteins called transcription factors, which are also created by the transcription-translation steps. Thus, there is an elaborate feedback mechanism to regulate protein levels in a cell as illustrated in Fig. 1. This networked system is called gene regulatory network.

In this paper, we consider the gene regulatory network, where each protein activates or represses another transcription in a cyclic way as depicted in Fig. 1. In particular, time delays arising from transportation and intermediate chemical reactions are explicitly considered. The dynamics of the cyclic gene regulatory network composed of $N$ genes is modeled as

$$\begin{cases} \dot{r}_i(t) = -a_i r_i(t) + \beta_i f_i(p_{i-1}(t - \tau_{p_{i-1}})), \\ \dot{p}_i(t) = -b_i p_i(t) + c_i r_i(t - \tau_{r_i}), \end{cases} \quad (1)$$

for $i = 1, 2, \cdots, N$, where $r_i, p_i \in \mathbb{R}_+ (:= \{x \in \mathbb{R} \mid x \geq 0\})$ denote the concentration of the $i$-th mRNA and its corresponding protein synthesized in the $i$-th gene, respectively (Chen and Aihara, 2002). For the sake of notational simplicity, we regard the subscript $0$ as $N$ and the subscript $N + 1$ as $1$ throughout this paper. Positive constants $a_i, b_i, c_i$ and $\beta_i$ have the following biological meanings: $a_i$ and $b_i$ denote the degradation rates of the $i$-th mRNA and protein, respectively: $c_i$ and $\beta_i$ denote the synthesis rates of the $i$-th mRNA and protein, respectively. A monotonic function $f_i : \mathbb{R}_+ \to \mathbb{R}_+$ represents either repression or activation of the transcription. For repression, $f_i(\cdot)$ is defined as $f_i(\cdot) := f^-(\cdot)$ such that $f^-(0) = 1$ and $f^-(\infty) = 0$. For activation, $f(\cdot) := f^+(\cdot)$ such that $f^+(0) = 0$ and $f^+(\infty) = 1$. In this paper, we use the following Hill functions:

$$f^-(p) := \frac{1}{1 + p^\nu}, \quad f^+(p) := \frac{p^\nu}{1 + p^\nu}, \quad (2)$$

where $\nu (\geq 1)$ is a Hill coefficient, which represents the degree of cooperative binding and determines the degree of nonlinearity of the system. The constants $\tau_{r_i}$ and $\tau_{p_i}$ ($i = 1, 2, \cdots, N$) represent time delays associated with the transcription and translation processes, respectively.

Let the following assumption be imposed throughout this paper:

**Assumption 1.**

$$\delta := \prod_{i=1}^{N} \delta_i < 0, \text{ where } \delta_i := \begin{cases} +1, & \text{for } f_i(\cdot) = f^+(\cdot), \\ -1, & \text{for } f_i(\cdot) = f^-(\cdot). \end{cases} \quad (3)$$

It is known that almost all solutions of (1) are observed to asymptotically converge to one of equilibria in the case of $\delta > 0$ (Mallet-Paret and Sell, 1996). Thus, it is reasonable to impose Assumption 1 to study the existence of oscillations, which is of our main interest in this paper.

## 2.2 Omega-limit set of the system

The omega-limit set of the gene regulatory network system (1) can be specified by using a Poincaré-Bendixson type theorem for time delay systems derived by Mallet-Paret and Sell (1996). The following proposition allows us to see that the omega-limit set of (1) is actually restricted to equilibrium points, periodic oscillations or homoclinic and heteroclinic orbits, and chaos is ruled out.

**Proposition 1.** (Mallet-Paret and Sell, 1996) *Consider the following system.*

$$\begin{aligned} \dot{x}_i(t) &= g_i(x_i(t), x_{i+1}(t)), \quad (i=1,2,\cdots,n-1) \\ \dot{x}_n(t) &= g_n(x_n(t), x_1(t-1)), \end{aligned} \quad (4)$$

*where $g_i(\cdot,\cdot)$ ($i = 1, 2, \cdots, n$) are $C^1$ nonlinear functions satisfying*

$$z_i \frac{\partial g_i(w,v)}{\partial v} > 0 \text{ and } z_i = \begin{cases} 1 & \text{if } i \neq n \\ z^* \in \{-1, 1\} & \text{if } i = n. \end{cases} \quad (5)$$

*Let $\boldsymbol{x}(t) = [x_1(t), x_2(t), \cdots, x_n(t)] \in \mathbb{R}^n$ be a solution of (4) on some interval $[t_0, \infty)$, and assume that $\boldsymbol{x}(t)$ is bounded in $\mathbb{R}^n$ as $t \to \infty$. Then, the omega-limit set of $\boldsymbol{x}(t)$ consists of*

(a) *a single non-constant periodic orbit,*

(b) *equilibrium points, or*

(c) *homoclinic and heteroclinic orbits.*

The dynamical model of gene regulatory networks (1) can be transformed to the form (4) satisfying (5) by letting $n = 2N$ and $x_i$ as follows.

$$\begin{cases} x_{2i-1}(t) = \sigma_{2i-1} p_{N-i+1}(Tt - \eta_{2i-1}), \\ x_{2i}(t) = \sigma_{2i} r_{N-i+1}(Tt - \eta_{2i}), \end{cases} \quad (6)$$

for $i = 1, 2, \cdots, N$, where

$$T := \sum_{j=1}^{2N} \tau_j \text{ and } \eta_i := \begin{cases} 0 & \text{for } i = 1 \\ \sum_{j=2}^{i} \tau_j & \text{for } i = 2, 3, \cdots, 2N \end{cases}$$

with $\tau_{2i-1} := \tau_{p_{N-i+1}}$ and $\tau_{2i} := \tau_{r_{N-i+1}}$. The constants $\sigma_i$ ($i = 1, 2, \cdots, 2N$) take either $+1$ or $-1$, and they are defined by

$$\sigma_i := \begin{cases} 1 & \text{for } i = 1 \\ \prod_{j=2}^{i} \rho_j & \text{for } i = 2, 3, \cdots, 2N, \end{cases}$$

where

$$\rho_{2i-1} := \text{sgn}\left[\frac{df_{N-i+2}}{dp}\right] \text{ for } i = 1, 2, \cdots, N.$$
$$\rho_{2i} := 1$$

The constant $z^*$ is then determined as $z^* = \prod_{i=1}^{2N} \rho_i$. The detailed proof is provided in Appendix A. Note that the above transformation affects only the sign of the omega-limit set, thus, the omega-limit set of $r_i(t)$ and $p_i(t)$ can be specified, once that of $x_i(t)$ is obtained.

Boundedness of $x(t)$ is easily verified in the similar way to El-Samad et al. (2005), where non-delay cyclic gene regulatory networks were considered. Existence of an equilibrium point and its uniqueness were proved in Hori et al. (2011) in the case of no time delay, and time delay does not affect these properties of the equilibrium point. Hence, the following lemma readily follows from Proposition 1.

**Lemma 1.** *Consider the cyclic gene regulatory networks modeled by (1). Then, the protein levels $p_i(t)$ ($i = 1, 2, \cdots, N$) exhibit (a) non-constant periodic oscillations, (b) convergence to the equilibrium point, or (c) homoclinic orbits.*

Note that chaos is ruled out for the system (1). This lemma immediately leads to the following proposition, which becomes a key to deriving existence conditions of oscillations in Section 4 and 5.

**Proposition 2.** *Consider the cyclic gene regulatory networks modeled by (1). The system has oscillations of protein levels $p_i(t)$ ($i = 1, 2, \cdots, N$), if the unique equilibrium point is locally unstable.*

In this paper, the term 'oscillations' refers to both non-constant periodic and homoclinic orbits. It should be noted that oscillations are periodic except for the case of homoclinic orbits.

## 3 Linearized Model of the Cyclic Gene Regulatory Networks

We see from Lemma 1 that the local instability at a unique equilibrium point implies the existence of oscillations. Thus, we here consider the linearized model of gene regulatory networks, and derive a unified formulation that is suitable for studying large-scale gene regulatory networks.

Consider a linearized system of (1) at the unique equilibrium point $[r_1^*, p_1^*, \cdots, r_N^*, p_N^*]^T$. Let $\hat{r}_i(s)$ and $\hat{p}_i(s)$ denote Laplace transform of $r_i(t)$ and $p_i(t)$, respectively. Then, Laplace transform of the linearized system with zero initial condition is obtained for $i = 1, 2, \cdots, N$ as

$$\begin{bmatrix} s\hat{r}_i(s) \\ s\hat{p}_i(s) \end{bmatrix} = \begin{bmatrix} -a_i & 0 \\ c_i e^{-s\tau_{r_i}} & -b_i \end{bmatrix} \begin{bmatrix} \hat{r}_i(s) \\ \hat{p}_i(s) \end{bmatrix} + \begin{bmatrix} \beta_i e^{-s\tau_{p_{i-1}}} \\ 0 \end{bmatrix} \hat{u}_i(s),$$

where

$$\hat{u}_i(s) := \zeta_i \hat{p}_{i-1}(s) \text{ and } \zeta_i := f_i'(p_{i-1}^*). \tag{7}$$

Thus, the transfer function of the $i$-th gene from $\hat{u}_i$ to $\hat{p}_i$ denoted by $h_i(s)$ is computed as

$$h_i(s) = \frac{R_i^2 e^{-s(\tau_{r_i} + \tau_{p_{i-1}})}}{(T_{r_i}s + 1)(T_{p_i}s + 1)}, \tag{8}$$

where

$$R_i := \frac{\sqrt{c_i \beta_i}}{\sqrt{a_i b_i}}, \quad T_{r_i} := \frac{1}{a_i}, \quad T_{p_i} := \frac{1}{b_i}. \tag{9}$$

The constant $R_i$ represents the ratio of the geometric means of synthesis rates and degradation rates of the $i$-th gene, and it is known as one of the essential biological quantities which characterize the oscillations in gene regulatory networks (Hori et al., 2011). The cyclic gene regulatory network system can be considered as the cyclic interconnection of the dynamical system $h_i(s)$ $(i = 1, 2, \cdots, N)$.

In order to capture the essential dynamical properties in an analytic way, we hereafter simplify the model so that the kinetic parameters $a_i, b_i, c_i$ and $\beta_i$ are homogeneous between genes.

**Assumption 2.** We assume $a_1 = a_2 = \cdots = a_N (=: a)$, and $b_1 = b_2 = \cdots = b_N (=: b)$, *i.e.*, mRNAs and proteins have common degradation rates between genes.

With the Assumption 2, the overall system can be written by a transfer function $\mathcal{H}(s)$ defined by

$$\mathcal{H}(s) := (\phi(s)e^{s\tau}I - K)^{-1}, \quad \phi(s) := \frac{1}{h(s)}, \tag{10}$$

where

$$h(s) := \frac{1}{(T_r s + 1)(T_p s + 1)}, \quad T_r := \frac{1}{a}, T_p := \frac{1}{b}, \tag{11}$$

$$\tau := \frac{1}{N} \sum_{i=1}^{N} (\tau_{r_i} + \tau_{p_i}), \tag{12}$$

$$K := \begin{bmatrix} 0 & 0 & \cdots & 0 & \zeta_1 R_1^2 \\ \zeta_2 R_2^2 & 0 & \cdots & 0 & 0 \\ 0 & \zeta_3 R_3^2 & \cdots & 0 & 0 \\ \vdots & \vdots & \ddots & \vdots & \vdots \\ 0 & 0 & \cdots & \zeta_N R_N^2 & 0 \end{bmatrix}. \tag{13}$$

The time delays $\tau_{r_i}$ and $\tau_{p_i}$ $(i = 1, 2, \cdots, N)$ of $h_i(s)$ can be different between genes, but the cyclic structure and the distributive property of linear systems allows us to equally distribute the time delays over all genes. Thus, the system $\mathcal{H}(s)$ shares the time delay $\tau$ among all genes, where $\tau$ represents the average of time delays in the network. Note that the structure of the matrix $K$ is determined from the graph topology of the gene regulatory network, and $\phi(s)e^{s\tau}$ is determined from the dynamics of each gene's expression.

In the next section, we will show that the instability of $\mathcal{H}(s)$ can be systematically checked using a simple graphical condition. Combined with Proposition 2, this graphical instability condition gives the existence condition of oscillations.

**Remark 1.** Although we impose Assumption 2 to develop qualitative rather than quantitative analysis framework, we can relax this assumption to some extent. In Appendix B, we discuss local instability of the unique equilibrium under the relaxed assumption. In particular, the homogeneous case, which is considered in the following sections, is shown to be a worst case for local instability under parameter perturbations. Thus, the analysis of the simplified model can also be interpreted as the worst-case analysis.

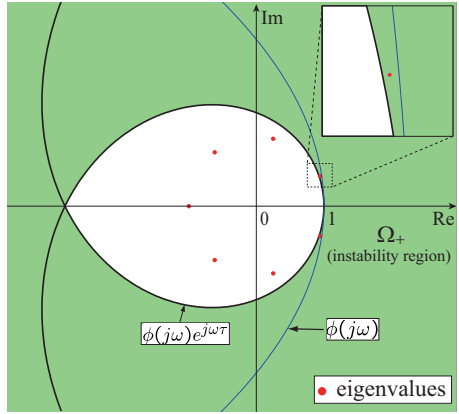

Figure 2: The instability region $\Omega_+$ and the eigenvalue locations of $K$.

# 4 Graphical Condition for the Existence of Oscillations

In this section, we present a graphical condition for the existence of oscillations. We first derive a necessary and sufficient graphical stability condition of $\mathcal{H}(s)$. Then, the existence condition is straightforwardly obtained, since the existence is guaranteed if $\mathcal{H}(s)$ is unstable as seen in Proposition 2.

## 4.1 Local stability condition

We first introduce an instability region of the large-scale linear system $\mathcal{H}(s)$, which is characterized by the gene's dynamics $h(s)e^{-s\tau}$. Let a set of complex values $\Omega_+$ be defined as

$$\Omega_+ := \{\lambda \in \mathbb{C} | \ \exists s \in \mathbb{C}_+, \ \phi(s)e^{s\tau} = \lambda\}, \tag{14}$$

where $\mathbb{C}_+ := \{s \in \mathbb{C} \mid \text{Re}[s] > 0\}$. The set $\Omega_+$ is the image of the open right-half complex plane under the mapping $\phi(s)e^{s\tau}$. The instability of $\mathcal{H}(s)$ can be characterized by $\Omega_+$ and the matrix $K$ as follows.

**Lemma 2.** *Consider the system $\mathcal{H}(s)$ defined by (10). Then, at least one pole of $\mathcal{H}(s)$ lies in the open right half plane of the complex plane, if and only if*

$$\text{spec}(K) \cap \Omega_+ \neq \emptyset. \tag{15}$$

The idea behind this lemma is essentially the same as the one in Proposition 5.1 of Hara et al. (2009). The complete proof is presented in Appendix C. An example of the instability region $\Omega_+$ and the eigenvalues of $K$ is illustrated in Fig. 2.

The stability counterpart of Lemma 2 is characterized by the set $\Omega_+^c := \{\lambda \in \mathbb{C} \mid \forall s \in \bar{\mathbb{C}}_+, \ \phi(s)e^{s\tau} \neq \lambda\}$ with $\bar{\mathbb{C}}_+ := \{s \in \mathbb{C} \mid \text{Re}[s] \geq 0\}$, which is an open complementary set of $\Omega_+$. Then, it follows that all the eigenvalues of $K$ lie in the stability region $\Omega_+^c$ if and only if $\mathcal{H}(s)$ is asymptotically stable.

**Remark 2.** Regarding the necessary and sufficient stability condition of $\mathcal{H}(s)$, Chen and Aihara (2002) presented a similar graphical test (see Theorem 2 in Chen and Aihara (2002)). The authors of this paper, however, have found that their graphical test is incorrect. We here briefly demonstrate a counterexample. More details are presented in Appendix D.

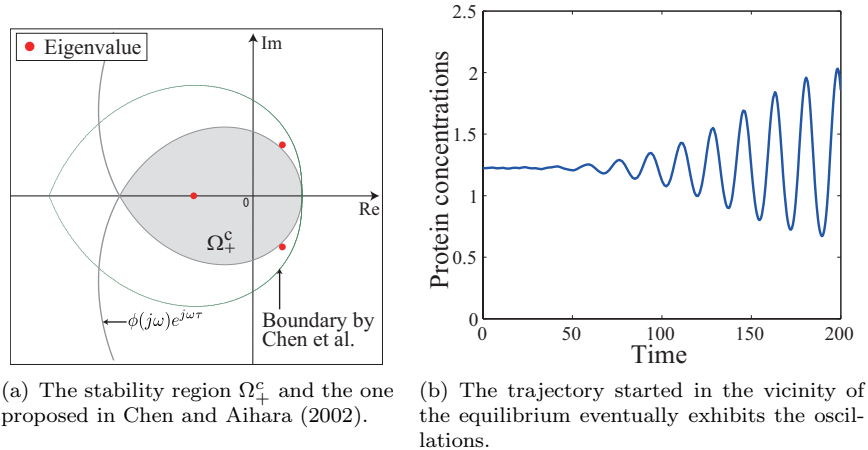

(a) The stability region $\Omega_+^c$ and the one proposed in Chen and Aihara (2002).

(b) The trajectory started in the vicinity of the equilibrium eventually exhibits the oscillations.

Figure 3: A counterexample to Theorem 2 in Chen and Aihara (2002).

Let $N = 3, a = b = 1.0000, c_i = \beta_i = 1.7498, \nu = 2.0000$ and $\tau_{r_i} = \tau_{p_i} = 0.5000$ ($i = 1, 2, 3$). It follows that $\phi(s)e^{s\tau} = (s+1)^2 e^s$, $R_1^2 = R_2^2 = R_3^2 = 1.7498$ and $\zeta_1 = \zeta_2 = \zeta_3 = 0.6858$. Theorem 2 in Chen and Aihara (2002) states that the system $\mathcal{H}(s)$ is stable if and only if all the eigenvalues of the matrix $K$ lie inside the region specified by an Archimedean spiral illustrated in Fig. 3(a). The eigenvalues of the matrix $K$ are obtained as $\{1.2e^{\pm j\pi/3}, -1.2\}$, and $\mathcal{H}(s)$ would be concluded as stable from Fig. 3(a).

However, the trajectory starting near the unique equilibrium exhibits oscillations as shown in Fig. 3(b), where the initial values are set as $[r_1, p_1, r_2, p_2, r_3, p_3] = [0.699, 1.224, 0.698, 1.226, 0.697, 1.225]$. In fact, we also show in Appendix D that the Nyquist contour of $\mathcal{H}(s)$ encloses $-1 + j0$, which implies that $\mathcal{H}(s)$ is unstable (see Fig. 8). This contradicts Theorem 2 in Chen and Aihara (2002). It should be noted that the stability condition presented in this paper concludes that $\mathcal{H}(s)$ is unstable as illustrated in Fig. 3(a).

## 4.2 A sufficient graphical condition

Recall that local instability of $\mathcal{H}(s)$ is a sufficient condition for the existence of oscillations as shown in Proposition 2, and the instability of $\mathcal{H}(s)$. The following graphical condition is the direct consequence of Proposition 2 and Lemma 2.

**Proposition 3.** *Consider the cyclic gene regulatory networks modeled by (1). Suppose Assumptions 1 and 2 hold. Then, the system has oscillations of protein concentrations $p_i(t)$ ($i = 1, 2, \cdots, N$), if*

$$\mathrm{spec}(K) \cap \Omega_+ \neq \emptyset. \tag{16}$$

Proposition 3 provides a graphical condition for the existence of oscillations. Given the dynamics of each gene $h(s)e^{-s\tau}$ and the interaction matrix $K$, the existence of oscillations can be characterized by the curve defined by the inverse of $h(s)e^{-s\tau}$ and the eigenvalues of $K$.

A remarkable feature of the graphical condition is that the eigenvalues are equally distributed on a circle with the center at the origin as illustrated in Fig. 2. Specifically, the eigenvalues of $K$ are written as

$$\lambda_k := L e^{\frac{j(2k-1)\pi}{N}} \quad (k = 1, 2, \cdots, N) \tag{17}$$

with

$$L := \prod_{i=1}^{N} |\zeta_i R_i^2|^{\frac{1}{N}}. \tag{18}$$

Note that $L$ is the radius of the circle. We designate $L$ as the average gain, since it is a geometric mean of the feedback gains of $K$ in (13).

Thus, Proposition 3 implies that the existence of oscillations is guaranteed if $\lambda_k \in \Omega_+$ for some $k = 1, 2, \cdots, N$. Moreover, the characteristic eigenvalue distribution allows us to simplify the graphical condition as follows.

**Theorem 1.** *Consider the cyclic gene regulatory networks modeled by (1). Suppose Assumptions 1 and 2 hold. Then, the system has oscillations of protein concentrations $p_i(t)$ $(i = 1, 2, \cdots, N)$, if $\lambda_k \in \Omega_+$ for some $k = 1, 2, \cdots, N$. Moreover, the following four conditions are equivalent.*

*(i) $\lambda_k \in \Omega_+$ for some $k = 1, 2, \cdots, N$.*

*(ii) $\lambda_1 \in \Omega_+$.*

*(iii) $\exists \omega_\sharp$ such that $\left|\phi(j\omega_\sharp)e^{j\omega_\sharp \tau}\right| < L$ and $\arg\left(\phi(j\omega_\sharp)e^{j\omega_\sharp \tau}\right) = \pi/N$.*

*(iv) $\exists \omega_*$ such that $\arg\left(\phi(j\omega_*)e^{j\omega_* \tau}\right) > \pi/N$ and $\left|\phi(j\omega_*)e^{j\omega_* \tau}\right| = L$.*

*where $\arg(\cdot)$ is the argument of a complex number.*

**Proof.** The condition (i) is equivalent to (16), thus the system (1) has oscillations if (i) is satisfied. We hereafter show the equivalence of the four conditions.
(i) $\Leftrightarrow$ (ii): The proof is mainly based on the fact that both $|\phi(j\omega)e^{j\omega t}|$ and $\arg(\phi(j\omega)e^{j\omega\tau})$ monotonically increase for positive $\omega$. The monotonicity is obvious from the definition (10). Then, it is easily verified that $\lambda_1$, which is the eigenvalue closest to the positive real axis, always goes inside the region $\Omega_+$ first, since the eigenvalues of the matrix $K$ are located on a circle center at the origin and radius $L$ (see Fig. 2).
(ii) $\Leftrightarrow$ (iii): Let $\omega_\sharp$ denote a frequency such that $\arg(\phi(j\omega_\sharp)e^{j\omega_\sharp \tau}) = \pi/N$. The conclusion immediately follows from the fact that $|\lambda_1| = L$ and $\arg(\lambda_1) = \pi/N$.
(iii) $\Leftrightarrow$ (iv): We only show (iii) $\Rightarrow$ (iv), since the converse can be shown in a similar manner. Suppose (iii) is satisfied. Let $\omega_*$ be defined such that $\left|\phi(j\omega_*)e^{j\omega_* \tau}\right| = L$. It follows that $\omega_\sharp < \omega_*$, because $|\phi(j\omega_\sharp)e^{j\omega_\sharp \tau}| < |\phi(j\omega_*)e^{j\omega_* \tau}| = L$ and there is the gain monotonicity for $\phi(j\omega)e^{j\omega\tau}$ as shown above. Then, the phase monotonicity implies $\arg(\phi(j\omega_\sharp)e^{j\omega_\sharp \tau}) = \pi/N < \arg(\phi(j\omega_*)e^{j\omega_* \tau})$. □

The condition (ii) in Theorem 1 greatly simplifies the graphical condition, because the existence of oscillations can be determined by the position of one specific eigenvalue $\lambda_1$ and the region $\Omega_+$. The condition (iii) is an analytic version of the consequence (ii), though it is generally difficult to obtain $\omega_\sharp$ in terms of the system's parameters. In the next section, the condition (iv) plays a key role to derive the analytic conditions for the existence of oscillations.

# 5 Analytic Condition for the Existence of Oscillations

## 5.1 Analytic conditions in terms of average gain

In this section, we provide analytic existence conditions of oscillations based on the geometric consideration of the graphical condition.

We first introduce normalized parameters of gene regulatory networks to avoid notational complexity, and to capture the essence of mathematical conditions. Let $T_A$ and $T_G$ denote the arithmetic and geometric means of the mRNA and protein degradation time constants, i.e.,

$$T_A := \frac{T_r + T_p}{2}, \quad T_G := \sqrt{T_r T_p}. \tag{19}$$

The constants $T_A$ and $T_G$ have the physical dimension of time. Define the following dimensionless constants $Q, \tilde{\omega}$ and $\tilde{\tau}$,

$$Q := \frac{T_G}{T_A}, \ \tilde{\omega} := \omega T_A, \ \tilde{\tau} := \frac{\tau}{T_A}. \tag{20}$$

Then, the boundary of the region $\Omega_+$ defined in (14) can be written as

$$\phi(j\omega)e^{j\omega\tau} = (-Q^2\tilde{\omega}^2 + 1 + 2j\tilde{\omega})e^{j\tilde{\omega}\tilde{\tau}}. \tag{21}$$

We see that the eigenvalues of $K$ and the region $\Omega_+$ are characterized in analytic form as (17) and (21), respectively. This leads to analytic conditions for the existence of oscillations. We first show existence conditions in terms of the average gain $L$ in (17).

**Theorem 2.** *Consider the gene regulatory networks modeled by (1). Suppose Assumptions 1 and 2 hold. Define the two functions $W(N,Q)$ and $D(Q,L)$ as*

$$W(N,Q) := \frac{2}{\cos\frac{\pi}{N} + \sqrt{\cos^2\frac{\pi}{N} + Q^2 \sin^2\frac{\pi}{N}}}, \tag{22}$$

$$D(Q,L) := 4(1-Q^2) + Q^4 L^2. \tag{23}$$

*Then, the system has oscillations of protein concentrations $p_i(t)$ $(i = 1, 2, \cdots, N)$, if one of the following two conditions holds* [1] *.*

(i) $L > W(N,Q)$,

(ii) $1 < L \leq W(N,Q)$ and

$$\arg\left(2 - \sqrt{D(Q,L)} + j2\sqrt{Q^2 - 2 + \sqrt{D(Q,L)}}\right)$$
$$> \frac{\pi}{N} - \frac{\sqrt{Q^2 - 2 + \sqrt{D(Q,L)}}}{Q^2}\tilde{\tau}. \tag{24}$$

**Proof.** It follows from Proposition 2 that there exists oscillations if the unique equilibrium point of (1) is unstable. Hence, we consider the instability condition of $\mathcal{H}(s)$, for which the simple graphical test in Lemma 2 is available.

We first consider the case of $L \leq 1$. It should be noted that the average gain $L$ is the radius of the circle where eigenvalues are located. It follows that $L \leq 1 \leq |\phi(j\omega)e^{j\omega\tau}|$ for all $\omega$. Since $|\phi(j\omega)e^{j\omega\tau}| = 1$ only when $\omega = 0$, and $\arg(\phi(j\omega)e^{j\omega\tau}) = 0$ for $\omega = 0$, there is no $\omega_*$ satisfying the condition (iv) in Theorem 1. Thus, Theorem 1 implies $\lambda_1 \notin \Omega_+$, and $\mathcal{H}(s)$ is not unstable.

---

[1] This condition is necessary and sufficient for local instability of $\mathcal{H}(s)$, which is readily seen from the proof.

In the case of $L > W(N, Q)$, we readily see $\lambda_1 \in \Omega_+$ according to Theorem 2 in Hori et al. (2011). In the case of $1 < L \leq W(N, Q)$, consider the condition (iv) in Theorem 1. Then, $\left|\phi(j\omega_*)e^{j\omega_*\tau}\right| = L$ yields

$$Q^4\tilde{\omega}_*^4 + 2(2 - Q^2)\tilde{\omega}_*^2 + 1 - L^2 = 0, \tag{25}$$

where $\tilde{\omega}_* := \omega_* T_A$. Then, $\tilde{\omega}_*$ is obtained as

$$\tilde{\omega}_* = \frac{\sqrt{Q^2 - 2 + \sqrt{D(Q, L)}}}{Q^2}, \tag{26}$$

and (24) is derived by substituting $\tilde{\omega}_*$ into $\arg(\phi(j\omega_*)e^{j\tilde{\omega}_*\tilde{\tau}}) > \pi/N$. □

The existence of oscillations can be determined by substituting the given parameters into Theorem 2. In particular, biological insights can be obtained by observing the relation between the quantities, which will be introduced in Section 6.2.

Since (24) has a certain monotone property in terms of $L$, we can simplify Theorem 2, and obtain the equivalent condition as follows.

**Corollary 1.** *Consider the cyclic gene regulatory networks modeled by (1). Suppose Assumptions 1 and 2 hold. Define $W(N, Q)$ and $D(Q, L)$ as (22) and (23), respectively. Then, the system has oscillations of protein concentrations $p_i(t)$ $(i = 1, 2, \cdots, N)$, if $L > \bar{L}$, where $\bar{L}$ is the solution of the equation*

$$\arg\left(2 - \sqrt{D(Q, \bar{L})} + j2\sqrt{Q^2 - 2 + \sqrt{D(Q, \bar{L})}}\right)$$
$$= \frac{\pi}{N} - \frac{\sqrt{Q^2 - 2 + \sqrt{D(Q, \bar{L})}}}{Q^2}\tilde{\tau}. \tag{27}$$

*In particular, the solution $\bar{L}$ is uniquely determined and satisfies $\bar{L} \in (1, W(N, Q)]$.*

The proof of Corollary 1 is provided in Appendix E.

**Remark 3.** In the case of $\tau = 0$, Theorem 2 and Corollary 1 coincide with Theorem 2 in Hori et al. (2011), which provides an existence condition of periodic oscillations for non-delay case.

## 5.2 Analytic conditions involving equilibrium point analysis

In Theorem 2 and Corollary 1, the value of $L$ depends on $\zeta_i$, which is determined from the equilibrium point $p_i^*$. Since the equilibrium point depends on the parameters $a, b, c_i$ and $\beta_i$, Theorem 2 and Corollary 1 require a numerical computation of the equilibrium point to determine $\zeta_i$. It is, however, desirable to explicitly take its dependence into account in the analytic conditions in order to gain biological insights on the relation between the parameters and the existence of oscillations. In this section, we restrict our attention to a class of the cyclic gene regulatory networks, and present analytic conditions for the existence of oscillations that explicitly consider the dependence of the equilibrium on the parameters.

Specifically, we consider the case where all interactions between genes are repressive, *i.e.*, $f_i(\cdot) = f^-(\cdot)$ for all $i = 1, 2, \cdots, N$, and $R_1 = R_2 = \cdots = R_N(=: R)$. It should be noted that this class of regulatory networks includes Repressilator (Elowitz and Leibler, 2000).

We first see that the equilibrium $p_i^*$ does not change by time delay, and $p_1^* = p_2^* = \cdots = p_N^*(= p^*)$ and $\zeta_1^* = \zeta_2^* = \cdots = \zeta_N^*(= \zeta^*)$ hold as shown in Hori et al. (2010). Using this property, we have the following relation between $L$ and $R$.

**Lemma 3.** *Consider the gene regulatory networks modeled by (1). Suppose $f_i(\cdot) = f^-(\cdot)$ ($i = 1, 2, \cdots, N$), $R_1 = R_2 = \cdots = R_N(=: R)$, and Assumptions 1 and 2 hold. Then $L < \nu$, and the following equality holds.*

$$R^2 = \left(\frac{L}{\nu - L}\right)^{1/\nu} \frac{\nu}{\nu - L}. \tag{28}$$

The proof of this lemma is provided in Appendix F.

Lemma 3 shows a direct relation between the average gain $L$ and the biological parameters, $R$ and $\nu$. Then, this lemma leads to the following existence condition of oscillations, which is explicitly written in terms of the biological parameters.

**Theorem 3.** *Consider the gene regulatory networks modeled by (1). Suppose $f_i(p) = f^-(p)$ ($i = 1, 2, \cdots, N$), $R_1 = R_2 = \cdots = R_N(= R)$, and Assumptions 1 and 2 hold. Define $W(N, Q), D(Q, L)$ and $\bar{L}$ as (22), (23) and (27), respectively. Then, the system has oscillations of protein concentrations $p_i(t)$ ($i = 1, 2, \cdots, N$) if both $\nu > \bar{L}$ and $R > \bar{R}$ hold, where*

$$\bar{R}^2 := \left(\frac{\bar{L}}{\nu - \bar{L}}\right)^{1/\nu} \frac{\nu}{\nu - \bar{L}}. \tag{29}$$

**Proof.** We derive an equivalent condition to Corollary 1. Observe that $R^2$ in (28) is monotonically increasing for $L(< \nu)$. Thus, if the condition $\bar{L} < L$ in Corollary 1 is satisfied, $\bar{R} < R$ follows, where $\bar{R}^2$ is defined by (29). We see from Lemma 3 that $\nu > \bar{L}$ is also satisfied, because $\nu > L$. On the other hand, if $\nu > \bar{L}$ and $R > \bar{R}$ are satisfied, we have $L > \bar{L}$ because of the monotonicity of (28).

Since the conditions $\nu > \bar{L}$ and $R > \bar{R}$ are equivalent to those of Corollary 1, we can conclude that there exists oscillations if these conditions are satisfied. □

Theorem 2 provides a condition for the existence of oscillations in terms of biological parameters $\nu, R$ and $\bar{R}(\nu, \bar{L}(N, Q, \tilde{\tau}))$ without any information about the equilibrium point $p^*$. This is contrast with the conditions in Theorem 2 and Corollary 1, where the dependence of $L$ on the equilibrium point $p^*$ is not explicitly obtained. Therefore, we can conclude from Theorem 2 that the following five dimensionless parameters characterize the existence of oscillations: the number of genes ($N$), time delay normalized by the arithmetic mean of the lifetime ($\tilde{\tau}$), the Hill coefficient ($\nu$), the ratio between the geometric mean of degradation rate and production rates ($R$), and the ratio between the geometric and arithmetic means of degradation rates ($Q$).

**Remark 4.** We can see that $\bar{R}$ and $\bar{L}$ are monotone with respect to the system's parameters. This observation leads to the conclusion that the system tends to have oscillations by increasing any of the five essential quantities, $N, \tilde{\tau}, \nu, R$ and $Q$. We will study more details in Section 6.2.

## 6 Numerical Example and Biological Insight

### 6.1 A cyclic gene regulatory network with $N = 7$

We here confirm the theoretical results provided in Theorems 1, 2 and 3 with illustrative numerical examples.

Consider the cyclic gene regulatory networks composed of $N = 7$ genes. Assume that $a = 1.2, b = 4.8, c_1 = c_3 = c_6 = c_7 = 1.92, c_2 = c_4 = c_5 = 3.84, \beta_1 = \beta_3 = \beta_6 = \beta_7 =$

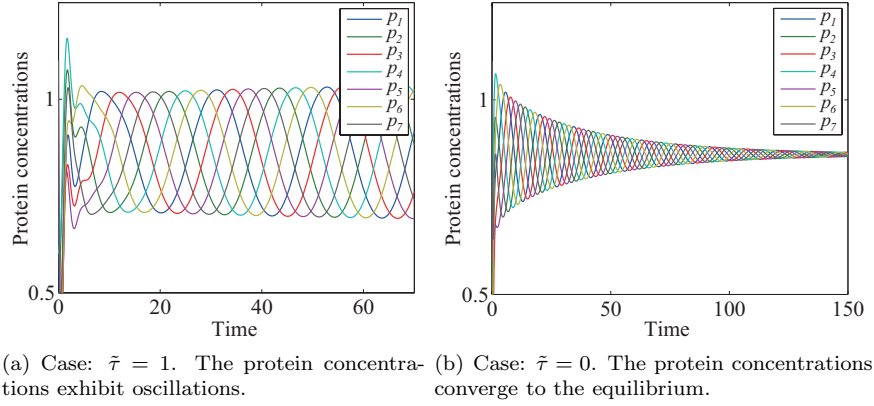

(a) Case: $\tilde{\tau} = 1$. The protein concentrations exhibit oscillations.

(b) Case: $\tilde{\tau} = 0$. The protein concentrations converge to the equilibrium.

Figure 4: Time plot of protein concentrations.

$4.32, \beta_2 = \beta_4 = \beta_5 = 2.16$, and let the Hill function be defined as $f_i(p) = 1/(1 + p^\nu)$ with $\nu = 2.6$ for $i = 1, 2, \cdots, N$. Then, $Q$ and $R(:= R_1 = R_2 = \cdots = R_7)$ are obtained as $Q = 0.800$ and $R = 1.200$ from the definition (20) and (9), respectively. The value of $L$ in (17) can be computed as $L = 1.048$. Note that $\zeta_i$ in $L$ involves computation of the unique equilibrium of the system, but it can be efficiently done with the bisection algorithm (see Hori et al. (2010) for details). In the following, we will see the effect of time delay by comparing a genetic regulatory network with and without time delay.

We first apply the graphical existence condition in Theorem 1. Figure 2 illustrates the instability region $\Omega_+$ and the eigenvalue distribution of $K$ for $\tilde{\tau} = 0$ and $\tilde{\tau} = 1.00$, where the time delays of the reactions are set as $\tau_{r_1} = \tau_{r_3} = \tau_{r_4} = \tau_{r_7} = 0.31, \tau_{r_2} = \tau_{r_5} = \tau_{r_6} = 0.26, \tau_{p_1} = \tau_{p_3} = \tau_{p_4} = \tau_{p_7} = 0.21, \tau_{p_2} = \tau_{p_5} = \tau_{p_6} = 0.26$, thus the average of the time delay is $\tau = 0.52$. In the case of $\tilde{\tau} = 1.00$, the boundary of the instability region $\Omega_+$ is given by $\phi(j\omega)e^{j\omega\tau}$ in Fig. 2. Thus, Theorem 1 implies the existence of oscillations, because two eigenvalues of $K$ belong to the region $\Omega_+$. In the case of $\tilde{\tau} = 0$, the boundary of the instability region $\Omega_+$ is $\phi(j\omega)$ in Fig. 2. We can see that all eigenvalues of $K$ are located outside the region $\Omega_+$ when $\tilde{\tau} = 0$. Thus, it is concluded from Theorem 1 that a unique equilibrium point of the system is locally asymptotically stable, and the protein concentrations do not exhibit oscillations when they are perturbed around the equilibrium point. Note that this result does not imply non-existence of oscillations, since Theorem 1 is a sufficient condition for the existence of oscillations.

The same conclusion follows from the analytic conditions in Theorem 2. We can see from Theorem 2 that there exist oscillations when $\tilde{\tau} = 1.00$, because $L = 1.048 > \bar{L} = 1.031$, where $\bar{L}$ is computed by (27). On the other hand, $L = 1.048 < \bar{L} = 1.072$ in the case of $\tilde{\tau} = 0$. Since the condition in Theorem 2 is equivalent to that of Theorem 1, we can conclude that the equilibrium point is locally asymptotically stable.

Theorem 2 and Corollary 1 required the value of equilibrium point to compute $L$. In contrast, Theorem 3 does not require the computation of the equilibrium. For given parameters, $\bar{L}$ and $\bar{R}$ can be determined from (27) and (29), respectively. Specifically, $\bar{L} = 1.031$ and $\bar{R} = 1.187$ when $\tilde{\tau} = 1.00$. Computing $R$ from (9) yields $R = 1.200$. Therefore, both of $\nu = 2.6 > \bar{L} = 1.031$ and $R = 1.200 > \bar{R} = 1.187$ in Theorem 3 are satisfied, and the existence of oscillations is concluded. On the other hand, $\bar{R} = 1.218$ when $\tilde{\tau} = 0$. Thus, the conditions in Theorem 3 do not hold, because $R = 1.200 \leq \bar{R} = 1.218$ despite $\nu = 2.6 > \bar{L} = 1.072$. It should be noted that this result, in turn, implies that there exists oscillations even for $\tilde{\tau} = 0$, if the parameters $a, b, c_i$ and $\beta_i$ are set so that $R > 1.218$ and $\bar{L} < 2.6$.

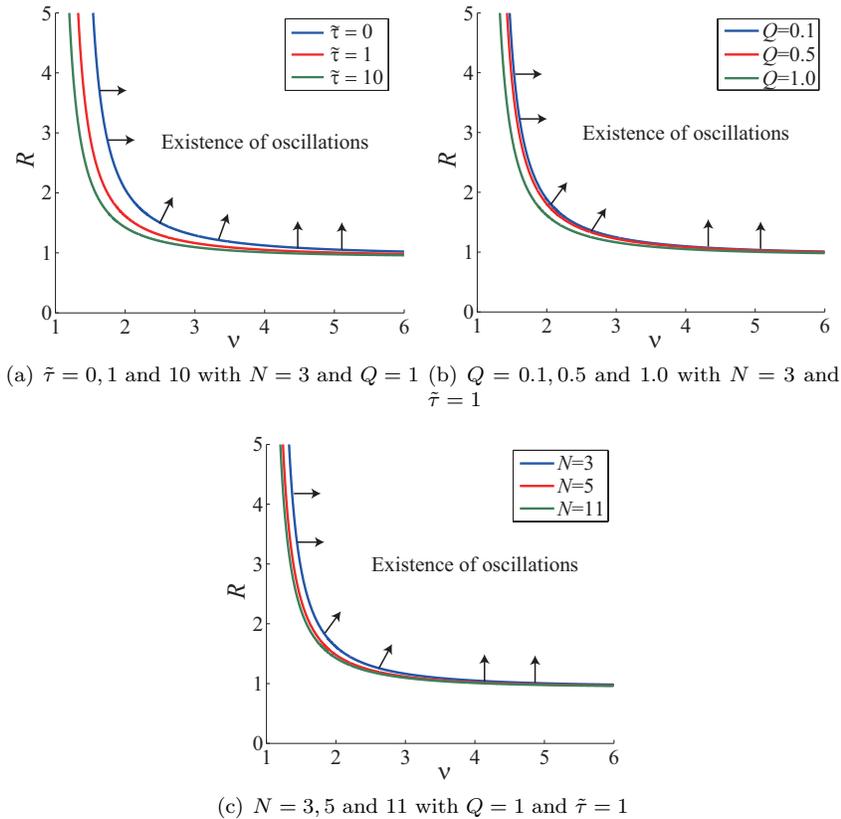

(a) $\tilde{\tau} = 0, 1$ and $10$ with $N = 3$ and $Q = 1$ (b) $Q = 0.1, 0.5$ and $1.0$ with $N = 3$ and $\tilde{\tau} = 1$

(c) $N = 3, 5$ and $11$ with $Q = 1$ and $\tilde{\tau} = 1$

Figure 5: Parameter regions $(\nu, R)$ for the existence of oscillations.

In fact, numerical simulations shown in Fig. 4 show oscillations and convergence to the equilibrium of protein concentrations for the time delay and non-delay case, respectively. We have seen that the existence of oscillations is more probable when the time delay is large. In what follows, we will see that this is indeed the case in general.

## 6.2 Biological insight: relation between parameters and existence of oscillations

As we have seen in Theorem 3, the existence of oscillations in cyclic gene regulatory networks with time delay can be characterized by the five dimensionless parameters $N, Q, \tilde{\tau}, \nu$ and $R$. The parameter regions that guarantee the existence of oscillations can be drawn as shown in Fig. 5 based on the analytic conditions given in Theorem 3. From these figures, we can readily conclude that the system tends to have oscillations as $\nu$ and $R$ get larger. In addition, the larger the parameters $\tilde{\tau}, Q$ and $N$ are, the more probable the existence of oscillations becomes because of each figure in Fig. 5.

An advantage of Theorem 3 is that we can confirm that these observations are true in general because the conditions are written in an analytic form in terms of the given biological parameters. Therefore, we conclude that the gene regulatory networks, in general, tends to have oscillations by letting the five essential parameters $N, Q, \tilde{\tau}, \nu$ and $R$ larger (see also Remark 4) [2].

---

[2] The constant $\bar{R}$ in Theorem 3 is not monotone decreasing for all $\nu$ ($\geq 1$), but it becomes a decreasing

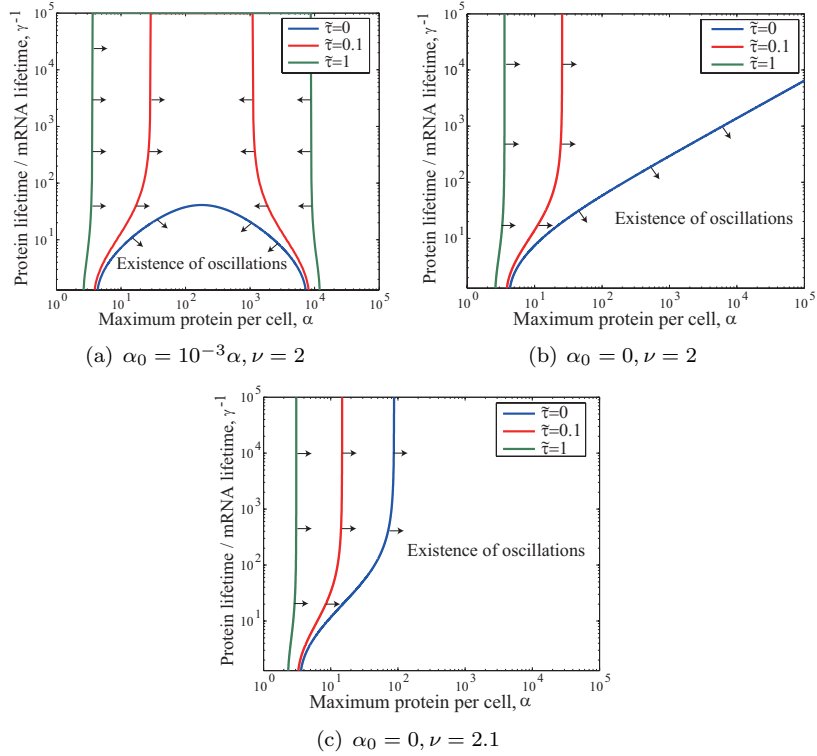

Figure 6: Parameter regions $(\alpha, \gamma^{-1})$ for the existence of oscillations in Repressilator. Both axes are in common log scale. Oscillations are more probable as the time delay becomes large.

# 7 Applications

In this section, we will apply our results to two existing biological systems, and see how our results work in analyzing the effect of time delay on the existence of oscillations.

## 7.1 Repressilator

Repressilator is one of the pioneering synthetic gene regulatory networks created by Elowitz and Leibler (2000). This artificial cyclic gene regulatory network is composed of three repressor genes, each of which represses another gene and forms cyclic reaction structure shown in Fig. 1. In Elowitz and Leibler (2000), Repressilator was implemented in *Escherichia Coli*, and oscillations of protein concentrations were observed in vitro.

The dynamical model of Repressilator can be written as

$$\begin{cases} \dot{r}_i(t) = -r_i(t) + \dfrac{\alpha}{1 + p_{i-1}(t - \tau_{r_{i-1}})^\nu} + \alpha_0, \\ \dot{p}_i(t) = -\gamma(p_i(t) - r_i(t - \tau_{r_i})), \end{cases} \qquad (30)$$

for $i = 1, 2, 3$, where $\gamma$ denotes the ratio of the protein degradation rate to the mRNA degradation rate, and the constant $\alpha_0$ represents leakiness of the promoter (Elowitz and Leibler, 2000). Note that time delays are not considered, *i.e.*, $\tau_{r_i} = \tau_{p_i} = 0$ for $i = 1, 2, 3$, in the original paper (Elowitz and Leibler, 2000). We remark that the recently proposed technique

---
function for $1 \leq \nu \leq 8$, which is the region of our interest.

(Ugander, 2008) could enable us to engineer the time delay, and it would contribute to obtaining a desired dynamics (see Orosz et al. (2010) for example). Hence, the time delays in the above model should account for such engineered delay as well as fast dynamics omitted in the modeling process. It can be seen that the model (30) is equivalent to (1) by rescaling the parameters when $\alpha_0 = 0$. We notice that Proposition 2 holds, even when $\alpha_0 \neq 0$, and Theorem 2 and Corollary 1 can be applied to analyze the existence of oscillations.

Let us first consider the case where no time delay appears in dynamics, *i.e.*, $\tau_{r_i} = \tau_{p_i} = 0$ $(i = 1, 2, \cdots, N)$, which is the original model of Repressilator presented in Elowitz and Leibler (2000). Following the numerical simulations conducted in Elowitz and Leibler (2000), we set $\alpha = 624, \alpha_0 = 0.0866, \beta = 0.200$ and $\nu = 2.0$. Then, $L$ and $\bar{L}$ in Corollary 1 can be computed as $L = 1.833$ and $\bar{L} = 1.519$, respectively. Thus, we conclude the existence of oscillations from Corollary 1, which is consistent with the simulation result in Elowitz and Leibler (2000).

Next, we investigate the effect of time delay on the existence of oscillations, and show that time delay increases robustness of Repressilator. Here we numerically examined the existence conditions in Theorem 2 for various time delays and parameters. The result is shown in Fig. 6, where the parameter regions for the existence of oscillations are illustrated. Note that only the normalized time delay $\tilde{\tau}$ rather than each time delay itself affects the existence of oscillations as seen in Section 6.2. Also note that the parameter region for $\tilde{\tau} = 0$ in Fig. 6 coincides with that in Fig. 1 (b) in Elowitz and Leibler (2000). We can see from Fig. 6 that the regions for the existence of oscillations get larger as $\tilde{\tau}$ become larger. This implies that one could make robust oscillator by inserting time delay. Moreover, the parameter region is not sensitive to a little change of $\alpha_0$ and $\nu$ when $\tilde{\tau}$ is large.

## 7.2 Somitogenesis clock

Somitogenesis is the process by which the somites of living organisms are created. Biological experiments as well as theoretical studies showed that the timing of the somite segmentation is regulated by an oscillatory expression of *Hes7* gene (see Hirata et al. (2004); Lewis (2003) and the references therein). In particular, it was shown by a biological experiment that oscillations produced by negative self-feedback of *Hes7* play a crucial role in controlling the somitegenesis oscillations (Hirata et al., 2004). In this section, we focus on the *Hes7* regulatory network, and see the validity of our theorems by comparing with the experimental data presented in Hirata et al. (2004). In addition, we provide some insights obtained from the developed theorems.

Following Hirata et al. (2004), we consider the following dynamical model of the regulatory network of the Hes7 protein.

$$\begin{cases} \dot{r}(t) = -ar(t) + \dfrac{\beta}{1 + (p(t - \tau_p)/p_0)^2}, \\ \dot{p}(t) = -bp(t) + cr(t - \tau_r). \end{cases} \quad (31)$$

This model is equivalent to the model (1) setting $N = 1$ and $\nu = 2$. Here, the mRNA and protein degradation rates $a$ and $b$ are defined as

$$a = \frac{\log 2}{t_r}, b = \frac{\log 2}{t_p}, \quad (32)$$

where $t_r$ and $t_p$ denote mRNA and protein half-life time. We employ the parameter values for wild-type Hes7 provided in Hirata et al. (2004): $t_p = 20$ [min], $t_r = 3$ [min], $a = 0.231$ [min$^{-1}$], $b = 0.0347$ [min$^{-1}$], $c = 4.5$ [min$^{-1}$], $\beta = 33$ [min$^{-1}$], $\tau_p = 30$ [min], $\tau_r = 7$ [min].

Table 1: Values of the essential biological parameters in somitogenesis oscillator

| Parameter | Before mutation | After mutation |
|---|---|---|
| $N$ | 1 | 1 |
| $Q$ | 0.674 | 0.575 |
| $\tilde{\tau}$ | 2.23 | 1.55 |
| $\nu$ | 2 | 2 |
| $R$ | 21.5 | 26.4 |
| $\bar{R}$ | 6.99 | - |
| $L$ | 1.97 | 1.97 |
| $\bar{L}$ | 1.85 | 2.39 |

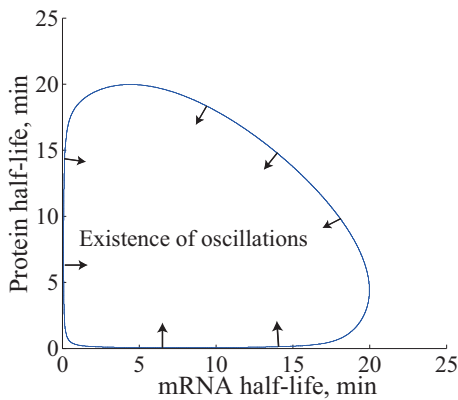

Figure 7: Parameter region for the existence of oscillations in somitogenesis oscillator. The existence of oscillations is more probable when mRNA half-life time is small, which is consistent with the hypothesis in Hirata et al. (2004).

In Hirata et al. (2004), a point mutation in the gene was introduced, and mice expressing mutant Hes7 were generated. The protein half-life of one of the Hes7 mutants was identified as almost $t_p = 30$ minutes, which is longer than that of the wild-type Hes7, which is $t_p = 20$ minutes. As a result, the protein degradation rate of the mutant Hes7 changed to $b = 0.0231$ [min$^{-1}$].

Numerical simulations of the model (31) revealed that the protein of the wild-type Hes7 shows oscillations, but that of the mutant Hes7 converges to a stable equilibrium (Hirata et al., 2004). In addition, the experimental result was consistent with the numerical simulations (Hirata et al., 2004).

We here present that Corollary 1 and Theorem 3 can explain these observations. First, we compute the values of the essential biological parameters, and obtain Table 1. We see from Table 1 that there exist oscillations before the mutation, because $L = 1.97 > \bar{L} = 1.85$ in Corollary 1, and equivalently $R = 21.5 > \bar{R} = 6.99$ in Theorem 3. On the other hand, the equilibrium point can be found to be locally stable after the mutation, because $L = 1.97 < \bar{L} = 2.39$ in Corollary 1, and $\bar{L} = 2.39 > \nu = 2$ in Theorem 3. We see that these results agree with the existing experimental work addressed above.

It was concluded in Hirata et al. (2004) that short half-life time $t_p$ of Hes7 protein is a key to the oscillations, though theoretical analysis was not performed. This hypothesis can be theoretically verified by using the presented theorems. Using Corollary 1, we can obtain the parameter region for the existence of oscillations in terms of half-life time of mRNA $t_r$ and protein $t_p$ in Fig. 7. We see that robust oscillations are guaranteed if the protein half-life time is shortened. For example, when mRNA half-life time is $t_r = 3$ [min], there exists oscillations for $0.1$ [min] $\leq t_p \leq 22$ [min].

## 8 Conclusion

In this paper, we have considered the conditions for the existence of oscillations in cyclic gene regulatory networks with time delay. Based on the unified analysis framework of gene regulatory network, we have first derived the graphical condition. Then, the geometric consideration has led to the analytic conditions. In particular, the relation between the equilibrium point and the parameters are explicitly considered, thus the condition is explicitly written in terms of biochemical parameters. Thus, biological insights can be easily obtained, and the relation between the parameter and the existence of oscillations has been revealed. Finally, we have confirmed that the developed theorems can be helpful to determined the existence of oscillations in two existing biological networks.

**Acknowledgments** This work was supported in part by the Ministry of Education, Culture, Sports, Science and Technology in Japan through Grant-in-Aid for Exploratory Research No. 19656104 and No. 21656106, Grant-in-Aid for Scientific Research (A) No. 21246067, and Grant-in-Aid for JSPS Fellows No. 23-9203.

## A  Transformation of the System

We here show that the gene regulatory network system defined by (1) can be obtained from (4) by the transformation (6).

We take a time derivative of $x_i(t)$ $(i = 1, 2, \cdots, 2N)$, and substitute (1).

$$\begin{aligned}
\dot{x}_{2i-1}(t) &= \sigma_{2i-1} T \dot{p}_{N-i+1}(Tt - \eta_{2i-1}) \\
&= \sigma_{2i-1} T \{-b_{N-i+1} p_{N-i+1}(Tt - \eta_{2i-1}) + \\
&\quad c_{N-i+1} r_{N-i+1}(Tt - \eta_{2i-1} - \tau_{r_{N-i+1}})\} \\
&= -b_{N-i+1} T x_{2i-1}(t) + \\
&\quad c_{N-i+1} T \sigma_{2i-1} r_{N-i+1}(Tt - \eta_{2i}) \\
&= -b_{N-i+1} T x_{2i-1}(t) + c_{N-i+1} T \rho_{2i} x_{2i}(t),
\end{aligned} \tag{33}$$

for $i = 1, 2, \cdots, N$.

$$\begin{aligned}
\dot{x}_{2i}(t) &= \sigma_{2i} T \dot{r}_{N-i+1}(Tt - \eta_{2i}) \\
&= \sigma_{2i} T \{-a_{N-i+1} r_{N-i+1}(Tt - \eta_{2i}) + \\
&\quad \beta_{N-i+1} f_{N-i+1}(p_{N-i}(Tt - \eta_{2i} - \tau_{p_{N-i}}))\} \\
&= -a_{N-i+1} T x_{2i}(t) + \\
&\quad \beta_{N-i+1} T \sigma_{2i} f_{N-i+1}(p_{N-i}(Tt - \eta_{2i+1})) \\
&= -a_{N-i+1} T x_{2i}(t) + \\
&\quad \beta_{N-i+1} T \sigma_{2i} f_{N-i+1}(\sigma_{2i+1} x_{2i+1}(t)),
\end{aligned} \tag{34}$$

for $i = 1, 2, \cdots, N-1$, and

$$\begin{aligned}
\dot{x}_{2N}(t) &= \sigma_{2N} T \dot{r}_1(Tt - \eta_{2N}) \\
&= \sigma_{2N} T \{-a_1 r_1(Tt - \eta_{2N} + \\
&\quad \beta_1 \sigma_{2N} f_1(p_N(Tt - \eta_{2N} - \tau_{p_N})))\} \\
&= -a_1 T x_{2N}(t) + \beta_1 T \sigma_{2N} f_1(p_N(Tt - T)) \\
&= -a_1 T x_{2N}(t) + \beta_1 T \sigma_{2N} f_1(x_1(t-1)).
\end{aligned} \quad (35)$$

We see that (33), (34) and (35) are of the form (1). Also we can verify that (33), (34) and (35) satisfy (5) as follows. It holds that

$$\frac{\partial g_{2i-1}(x_{2i-1}, x_{2i})}{\partial x_{2i}} = c_{N-i+1} T \rho_{2i}, \quad (36)$$

$$\frac{\partial g_{2i}(x_{2i}, x_{2i+1})}{\partial x_{2i+1}} = \beta_{N-i+1} T \sigma_{2i} \sigma_{2i+1} \frac{df_{N-i+1}}{dp}, \quad (37)$$

$$\frac{\partial g_{2N}(x_{2N}, x_1)}{\partial x_1} = \beta_1 T \sigma_{2N} \frac{df_1}{dp}. \quad (38)$$

It is clear from the definition that (36) is positive. We can see that (37) and (38) are also positive, because it follows that

$$\sigma_{2i} \sigma_{2i+1} \mathrm{sgn}\left[\frac{df_{N-i+1}}{dp}\right] = \rho_{2i+1}^2 > 0, \sigma_{2N} \mathrm{sgn}\left[\frac{df_1}{dp}\right] = z^*.$$

Note that the sign of $x_i$ is the same as that of $\sigma_i$, and $f_i(\cdot)$ is a monotonic function defined on positive orthant. The relation (5) can be alternatively confirmed by the following calculation.

$$\begin{aligned}
\frac{\partial g_{2i-1}(x_{2i-1}, x_{2i})}{\partial x_{2i}} &= \frac{\partial r_{N-i+1}}{\partial x_{2i}} \frac{\partial g_{2i-1}}{\partial r_{N-i+1}} = \sigma_{2i} c_{N-i+1} T \sigma_{2i-1} \\
&= c_{N-i+1} T \rho_{2i}, \\
\frac{\partial g_{2i}(x_{2i}, x_{2i+1})}{\partial x_{2i+1}} &= \frac{\partial p_{N-i}}{\partial x_{2i+1}} \frac{\partial g_{2i}}{\partial p_{N-i}} \\
&= \sigma_{2i+1} \beta_{N-i+1} T \sigma_{2i} \frac{df_{N-i+1}}{dp}, \\
\frac{\partial g_{2N}(x_{2N}, x_1)}{\partial x_1} &= \frac{\partial p_N}{\partial x_1} \frac{\partial g_{2N}}{\partial p_N} = \sigma_1 \beta_1 T \sigma_{2N} \frac{df_1}{dp} \\
&= \beta_1 T \sigma_{2N} \frac{df_1}{dp},
\end{aligned}$$

where the right-hand sides coincide with those of (36), (37) and (38), respectively.

## B  Discussions on Robustness of Local Instability

In this section, we relax Assumption 2 and discuss local instability of the unique equilibrium of (1) (see also Remark 1). Specifically, we consider robust instability of the linearized system under parameter perturbations, which are often the case in biochemical systems. Let $\mathcal{P}$ denote a set of parameters defined by

$$\mathcal{P} := \{(a_i, b_i, c_i, \beta_i, \tau_{r_i}, \tau_{p_i}, \zeta_i) \ (i = 1, 2, \cdots, N) \mid \underline{a_i} \le a_i \le \overline{a_i}, \underline{b_i} \le b_i \le \overline{b_i}, \underline{c_i} \le c_i \le \overline{c_i},$$
$$\underline{\beta_i} \le \beta_i \le \overline{\beta_i}, \underline{\tau_{r_i}} \le \tau_{r_i} \le \overline{\tau_{r_i}}, \underline{\tau_{p_i}} \le \tau_{p_i} \le \overline{\tau_{p_i}}, \underline{\zeta_i} \le |\zeta_i| \le \overline{\zeta_i} \ (i = 1, 2, \cdots, N)\}, \quad (39)$$

where the symbols with a underline and an overline, *i.e.,* $\underline{a_i}$ and $\overline{a_i}$ etc., represent given upper and lower bounds of the parameters, respectively. The perturbation of the linearized gain $\zeta_i$ $(i = 1, 2, \cdots, N)$ accounts for both the uncertainty of equilibrium due to the perturbations of $(a_i, b_i, c_i, \beta_i)$ and that of the Hill coefficient $\nu$. We define the linearized system by $\tilde{\mathcal{H}}(s) := (I - H(s)M)^{-1}$, where $H(s) := \text{diag}(h_1(s), h_2(s), \cdots, h_N(s))$ and

$$M := \begin{bmatrix} 0 & 0 & \cdots & 0 & \zeta_1 \\ \zeta_2 & 0 & \cdots & 0 & 0 \\ 0 & \zeta_3 & \cdots & 0 & 0 \\ \vdots & \vdots & \ddots & \vdots & \vdots \\ 0 & 0 & \cdots & \zeta_N & 0 \end{bmatrix}. \tag{40}$$

In Hori et al. (2011), it was shown for $\tau_{r_i} = \tau_{p_i} = 0$ that the worst-case perturbation for local instability is given by $(a_i, b_i, c_i, \beta_i, \zeta_i) = (\overline{a_i}, \overline{b_i}, \underline{c_i}, \underline{\beta_i}, \delta_i \underline{\zeta_i})$, which is the upper and lower extremum of the parameter set. We can easily obtain a similar result for $\tau_{r_i} \neq 0$ and $\tau_{p_i} \neq 0$.

**Proposition 4.** *Consider the cyclic gene regulatory networks modeled by (1). Suppose Assumption 1 holds. Then, the linearized system $\tilde{\mathcal{H}}(s)$ is unstable for all parameters in $\mathcal{P}$, if and only if $\tilde{\mathcal{H}}(s)$ with*

$$a_i = \overline{a_i}, b_i = \overline{b_i}, c_i = \underline{c_i}, \beta_i = \underline{\beta_i}, \tau_{r_i} = \underline{\tau_{r_i}}, \tau_{p_i} = \underline{\tau_{p_i}}, \zeta_i = \delta_i \underline{\zeta_i} \tag{41}$$

*is unstable.*

**Proof.** The idea of the proof is essentially the same as the one presented in Hori et al. (2011), but we here provide the detailed proof for completeness.

The poles of $\tilde{\mathcal{H}}(s)$ are given by the roots of

$$|I - H(s)M| = 1 + \prod_{i=1}^{N} h_i(s)|\zeta_i| = 0 \iff \frac{e^{s(\tau_{r_1} + \tau_{p_N})}}{c_1 \beta_1}(s + a_1)\gamma(s) + \prod_{i=1}^{N}|\zeta_i| = 0, \tag{42}$$

where $\gamma(s) := (s + b_1) \prod_{i=1}^{N}(1/h_i(s))$. We first fix the parameters $a_i (i = 2, 3, \cdots, N)$ and $b_i, c_i, \beta_i, \tau_{r_i}, \tau_{p_i}$ $(i = 1, 2, \cdots, N)$. Let $a_\rho (= a_1)$ and $\omega_\rho (= \omega)$ satisfy (42), which implies

$$v_\rho^* = \frac{1}{c_1 \beta_1}\sqrt{a_\rho^2 + \omega_\rho^2}|\gamma(j\omega_\rho)| \text{ and } \angle(a_\rho + j\omega_\rho) + \omega_\rho(\tau_{r_1} + \tau_{p_N}) = \pi - \angle\gamma(j\omega_\rho), \tag{43}$$

where $v_\rho^* := \prod_{i=1}^{N}|\zeta_i|$ is the critical gain for destabilizing the closed-loop system.

Here, we perturb $a_\rho$ to $a_\nu (> a_\rho)$. Then, it follows that

$$\angle(a_\nu + j\omega_\rho) + \omega_\rho(\tau_{r_1} + \tau_{p_N}) < \angle(a_\nu + j\omega_\rho) + \omega_\rho(\tau_{r_1} + \tau_{p_N}) = \pi - \angle\gamma(j\omega_\rho). \tag{44}$$

Moreover, there exists $\omega_\nu (> \omega_\rho)$ such that $\angle(a_\nu + j\omega_\nu) + \omega_\nu(\tau_{r_1} + \tau_{p_N}) = \pi - \angle\gamma(j\omega_\nu)$, since $\gamma(j\omega)$ is an increasing function in terms of $\omega$. Then, the critical gain $v_\nu^*$ that destabilizes the closed-loop system is given by

$$v_\nu^* = \frac{1}{c_1 \beta_1}\sqrt{a_\nu^2 + \omega_\nu^2}|\gamma(j\omega_\nu)| > v_\rho^*, \tag{45}$$

because $\omega_\nu > \omega_\rho$ and $|\gamma(j\omega)|$ is an increasing function for $\omega > 0$.

Thus, the critical gain, which destabilizes the closed-loop system $\tilde{\mathcal{H}}(s)$, monotonically increases as $a_1$ increases. This implies that $\tilde{\mathcal{H}}(s)$ is unstable for all $a_1 \in [\underline{a_1}, \overline{a_1}]$, if and only if the system with $a_1 = \overline{a_1}$ is unstable. The same conclusion follows for $a_i$ $(i = 2, 3, \cdots, N)$ and $b_i (i = 1, 2, \cdots, N)$ by the same proof.

Regarding $c_i$ and $\beta_i$ ($i = 1, 2, \cdots, N$), we see that only the gain condition in (43), which is the first equality, depends on $c_i$ and $\beta_i$. Thus, we can immediately conclude that $c_i = \underline{c_i}$ and $\beta_i = \underline{\beta_i}$ are the worst-case parameters for instability of $\tilde{\mathcal{H}}(s)$.

We can apply the same discussion to the delays. Suppose $\tau_\rho (= \tau_{r_1})$ satisfy (43). We define a perturbed delay as $\tau_\nu (> \tau_\rho)$. Then, $\angle(a_1 + j\omega_\rho) + \omega_\rho(\tau_\nu + \tau_{p_N}) > \pi - \gamma(j\omega_\rho)$, and we see that there exists $\omega_\nu < \omega_\rho$ such that $\angle(a_1 + j\omega_\nu) + \omega_\nu(\tau_\nu + \tau_{p_N}) = \pi - \gamma(j\omega_\nu)$. Thus, the critical gain for instability decreases as $\tau_{r_1}$ increases. This, in turn, means that $\tau_{r_1}$ is the worst-case delay for robust instability. Applying the same proof for $\tau_{r_i}$ ($i = 2, 3, \cdots, N$) and $\tau_{p_i}$ ($i = 1, 2, \cdots, N$), we have the conclusion. □

This proposition means that we can guarantee the existence of periodic oscillations for all parameters in $\mathcal{P}$ only from the local instability analysis for the extreme parameters.

Let $\overline{a} := \max_i \overline{a_i}$, $\overline{b} := \max_i \overline{b_i}$, $\underline{c} := \min_i \underline{c_i}$, $\underline{\beta} := \min_i \underline{\beta_i}$, $\underline{\tau_r} := \min_i \underline{\tau_{r_i}}$ and $\underline{\tau_p} := \min_i \underline{\tau_{p_i}}$. Proposition 4 immediately leads to the following sufficient condition for local instability.

**Corollary 2.** *Consider the cyclic gene regulatory networks modeled by (1). Suppose Assumption 1 holds. Then, the linearized system $\tilde{\mathcal{H}}(s)$ is unstable for all parameters in $\mathcal{P}$, if $\mathcal{H}(s)$, which is defined in (10), is unstable for*

$$a = \overline{a}, b = \overline{b}, c_i = \underline{c}, \beta_i = \underline{\beta}, \tau_{r_i} = \underline{\tau_r}, \tau_{p_i} = \underline{\tau_p}, \text{ and } \zeta_i = \delta_i \underline{\zeta_i}. \tag{46}$$

This corollary means that the equilibrium point is locally unstable for all $\mathcal{P}$, if the homogeneous system $\mathcal{H}(s)$ defined by (10) is unstable for the extreme parameters (46). Note that local instability of the equilibrium is a sufficient condition for the existence of oscillations as shown in Proposition 2. Although Assumption 2 is made in order primarily to simplify the model and gain qualitative insights, Corollary 2 implies that the analysis of the simplified model also provides conditions for the robustness of the existence oscillations.

## C  Proof of Lemma 2

The poles of $\mathcal{H}(s)$ are obtained by solving

$$|\phi(s)e^{s\tau} I - K| = |\phi(s)e^{s\tau} I - \Lambda| = 0, \tag{47}$$

where $\Lambda := \mathrm{diag}(\lambda_1, \lambda_2, \cdots, \lambda_N) \in \mathbb{C}^{N \times N}$. It should be noted that $\mathcal{H}(s)$ is a retarded time delay system, thus the dominant pole is located at the rightmost position in the complex plane. Thus, it follows from the definition of $\Omega_+$ that $\mathcal{H}(s)$ has at least one pole in $\mathbb{C}_+$ if and only if $\mathrm{spec}(K) \cap \Omega_+ \neq \emptyset$. □

## D  A Counterexample to Theorem 2 in Chen and Aihara (2002)

In this section, we first examine the stability of $\mathcal{H}(s)$ by numerically computing the Nyquist plot and the poles of $\mathcal{H}(s)$. We then point out the errors of the proof in Chen and Aihara (2002).

Let the parameters of (1) be set as the ones in Remark 2. It follows that $\phi(s)e^{s\tau} = (s+1)^2 e^s$, $R_1^2 = R_2^2 = R_3^2 = 1.7498$ and $\zeta_1 = \zeta_2 = \zeta_3 = 0.6858$. In what follows, we show that $\mathcal{H}(s)$ is actually unstable, though Theorem 2 in Chen and Aihara (2002) concludes that it is stable as explained in Remark 2.

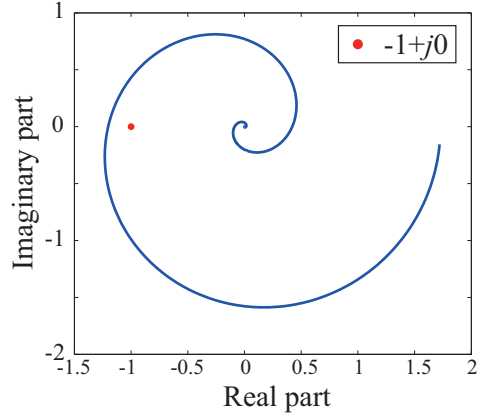

Figure 8: Nyquist contour of the system. The contour encloses -1+j0.

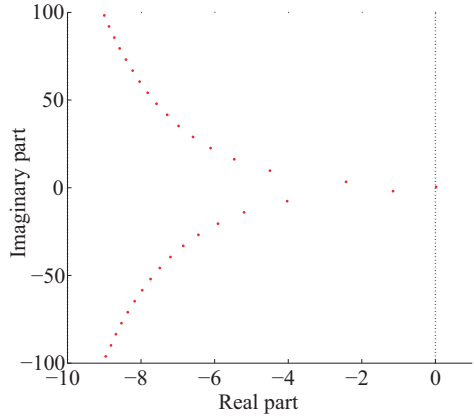

Figure 9: Roots of the characteristic equation in terms of the eigenvalue $\sqrt{1.2}e^{j\pi/6}$.

Figure 8 illustrates the Nyquist plot of the loop transfer function of $\mathcal{H}(s)$, which is defined by

$$-\prod_{i=1}^{3}\frac{c_i\beta_i}{(s+a_i)(s+b_i)}e^{-s(\tau_{r_i}+\tau_{p_i})}\zeta_i = 1.2^3\frac{e^{-3s}}{(s+1)^6}.$$

We see that the Nyquist contour encloses $-1+j0$. Since the open loop system is stable, the Nyquist contour in Fig. 8 implies that the system is unstable, which contradicts Theorem 2 in Chen and Aihara (2002). In fact, a pair of the poles of $\mathcal{H}(s)$ is found in the open right-half complex plane at $0.0212 \pm 0.3634j$ by numerical computation (see Fig. 9). These observations imply that Theorem 2 in Chen and Aihara (2002) is not the necessary and sufficient stability condition for $\mathcal{H}(s)$.

We hereafter clarify the errors of their mathematical proof. There are essentially two errors in the mathematical proof provided in Chen and Aihara (2002). First, Theorem 2.6 in Belair (1993), which is used in the proof in Chen and Aihara (2002), is incorrect. Second, Theorem 2.6 in Belair (1993) was applied in a wrong way in Chen and Aihara (2002).

In the remaining of this section, we use the notations defined in Belair (1993); Chen and Aihara (2002) for the sake of easy comprehension and comparison. Our first claim is that Theorem 2.6 in Belair (1993) is not the necessary and sufficient, but a sufficient condition. Using the notations in Belair (1993), we see that

$$\lambda = -1 + be^{-\lambda\tau} \iff \sigma e^{\sigma\tau} = be^{\tau}$$
$$\iff \begin{cases} R = \rho e^{(r-1)\tau} \\ \phi = \theta + s\tau + 2\pi j \end{cases}$$
$$\iff \begin{cases} R = \rho e^{(r-1)\tau} & \text{(a)} \\ \phi = \theta + Re^{(1-r)\tau}\tau\sin\theta & \text{(b)}. \end{cases}$$

where $\rho > 0$ and $0 \leq \theta < 2\pi$ [3]. Note that the equations (a) and (b) are the same as (2.7a) and (2.8) in Belair (1993). It was concluded that all roots $\lambda$ of the above equation have negative real parts if and only if $(R, \phi) \in \{(R, \phi) \in [0, \infty) \times [0, 2\pi) \mid \text{(a) and (b) only if } r < 1\}$. Then, the stability region was considered by specifying $(R, \phi)$ that belongs to the above set. It follows that

$$\{(R, \phi) \in [0, \infty) \times [0, 2\pi) \mid \text{(a) and (b) only if } r < 1\} \tag{48}$$
$$\supsetneq \{(R, \phi) \in [0, \infty) \times [0, 2\pi) \mid \text{(b) only if } r < 1\} \tag{49}$$
$$= \overline{\{(R, \phi) \in [0, \infty) \times [0, 2\pi) \mid \text{(b) for some } r \geq 1\}}, \tag{50}$$

where $\overline{\{\cdot\}}$ denotes a complementary set. Then, the set (50) was specified in Theorem 2.6 in Belair (1993) as

$$\overline{\{(R, \phi) \in [0, \infty) \times [0, 2\pi) \mid \text{(b) for some } r \geq 1\}},$$
$$= \overline{\left\{(R, \phi) \in [0, \infty) \times [0, 2\pi) \mid \phi \leq \frac{\pi}{2} + R\tau \text{ or } \phi \geq \frac{3\pi}{2} - R\tau\right\}}$$
$$= \left\{(R, \phi) \in [0, \infty) \times [0, 2\pi) \mid \phi > \frac{\pi}{2} + R\tau \text{ and } \phi < \frac{3\pi}{2} - R\tau\right\}. \tag{51}$$

In Belair (1993), however, (49) was not derived as a subset but as an equivalent set of (48). Therefore, it was concluded that (51) provides the stability region where all the eigenvalues of the system are located, *if and only if* the system is asymptotically stable. This conclusion is, however, incorrect, because (49) is actually a subset of (48).

Instead, we see that (51) is the region where all the eigenvalues of the system are located, *if* the system is asymptotically stable. Therefore, we claim that Theorem 2.6 in Belair (1993) provides only a sufficient condition for stability.

The other error of Theorem 2 in Chen and Aihara (2002) stems from the mis-application of Theorem 2.6 in Belair (1993). The boundary of the stability region provided in Theorem 2.6 in Belair (1993) is the Archimedean spiral starting from the origin. Applying Theorem 2.6 in Belair (1993) to the equation (11) in Chen and Aihara (2002), we see that the boundary of the stability region is given by

$$R = \begin{cases} \dfrac{2\theta - \pi}{k\tau} & \text{for } \dfrac{\pi}{2} < \theta < \pi \\ \dfrac{3\pi - 2\theta}{k\tau} & \text{for } \pi < \theta < \dfrac{3\pi}{2}, \end{cases} \tag{52}$$

where the constants $k$ and $\tau$ are defined as in Chen and Aihara (2002). The tuple $(R, \theta)$ defines the distance and the angle of the boundary measured from the origin. Consequently, the arc drawn by (52) becomes the well-known Archimedean spiral.

---

[3] Equation (2.4) in Belair (1993) is typo. It should be corrected as $\lambda = -1 + be^{-\lambda\tau}$.

In Theorem 2 of Chen and Aihara (2002), however, the left-hand side of (52), $R$, was shifted by one, and the equation of the boundary was given by

$$R - 1 = \begin{cases} \dfrac{2\theta - \pi}{k\tau} & \text{for } \dfrac{\pi}{2} < \theta < \pi \\ \dfrac{3\pi - 2\theta}{k\tau} & \text{for } \pi < \theta < \dfrac{3\pi}{2}. \end{cases} \tag{53}$$

Then, $R$ and $\theta$ are measured from the origin and $1 + j0$, respectively (see also Fig. 2 in Chen and Aihara (2002)). It is clear that the boundary obtained in this way does not coincide with the one in Theorem 2.6 in Belair (1993).

## E  Proof of Corollary 1

We observe that the left-hand side of (24) is the monotonically increasing function of $L$, and the right-hand side is the monotonically decreasing function of $L$. Moreover, we can see that the inequality (24) is not satisfied when $L \leq 1$, but it is satisfied when $L > W(N, Q)$. Therefore, we have a critical value $\bar{L}$ at which the left-hand side and the right-hand side of (24) take the same value, and the inequality (24) is satisfied if and only if $L > \bar{L}$. It is clear from the above argument that $\bar{L}$ is given as the unique solution of (27), and $\bar{L}$ satisfies $1 < \bar{L} \leq W(N, Q)$.

## F  Proof of Lemma 3

It follows that

$$L = -R^2 \zeta = -R^2 \left( \frac{-\nu p^{*\nu-1}}{(1 + p^{*\nu})^2} \right), \tag{54}$$

where the second equality follows from the definition of $\zeta$. According to Hori et al. (2010), we have

$$p^* = \frac{R^2}{1 + p^{*\nu}}. \tag{55}$$

Then, it follows from (54) and (55) that

$$L = \frac{\nu p^{*\nu}}{1 + p^{*\nu}}. \tag{56}$$

This implies $L < \nu$. In addition, it follows that

$$\zeta = -\frac{\nu}{R^4}(R^2 - p^*) \tag{57}$$

(see Hori et al. (2010) for the details). Multiplicating $R^2$ to (57), we have

$$L = -\frac{\nu}{R^2}(R^2 - p^*). \tag{58}$$

Thus, we can eliminate $p^*$ from (56) by using (58), and obtain (28).